\documentclass[a4paper]{article}

\usepackage{INTERSPEECH2015}

\usepackage{graphicx}
\usepackage{amssymb,amsmath,bm}
\usepackage{textcomp}
\usepackage{url}

\title{Significance of Maximum Spectral Amplitude in Sub-bands for Spectral Envelope Estimation and Its Application to Statistical Parametric Speech Synthesis}

\makeatletter
\def\name#1{\gdef\@name{#1\\}}
\makeatother \name{{\em Sivanand Achanta, Anandaswarup Vadapalli, Sai Krishna R, Suryakanth V Gangashetty}}

\address{International Institute of Information Technology, Hyderabad, INDIA \\
  {\small \tt \{sivanand.a,anandaswarup.vadapalli,saikrishna.r\}@research.iiit.ac.in, svg@iiit.ac.in}
  }

\begin{document}

\maketitle
\begin{abstract}
In this paper we propose a technique for spectral envelope estimation using maximum values in the sub-bands of Fourier magnitude spectrum (MSASB). Most other methods in the literature parametrize spectral envelope in cepstral domain such as Mel-generalized cepstrum etc. Such cepstral domain representations, although compact, are not readily interpretable.  This difficulty is overcome by our method which parametrizes in the spectral domain itself. In our experiments, spectral envelope estimated using MSASB method was incorporated in the STRAIGHT vocoder. Both objective and subjective results of analysis-by-synthesis indicate that the proposed method is comparable to STRAIGHT. We also evaluate the effectiveness of the proposed parametrization in a statistical parametric speech synthesis framework using deep neural networks.
\end{abstract}

\section{Introduction}
Statistical parametric speech synthesis (SPSS) has evolved as a parallel technique to unit selection for text-to-speech conversion. There are many reasons for the success of SPSS paradigm, one of them is being a small foot print system \cite{smfpspss} and the other being its flexibility to synthesize expressive voices \cite{expressivespss}. However, there are three key problems in the SPSS that many researchers have been attempting to tackle with \cite{sps}. First, the problem of vocoding i.e., the simple excitation model often used in most vocoders results in a buzzy synthesis. Second, the problem of acoustic modeling, the HMM-GMM model usually employed has only limited ability to model the dependencies across features in a speech frame. Third, the problem of over smoothing resulting because of the way parameters are generated from the acoustic model. In this paper, we address the issue of spectrum parametrization relating to the first of these problems namely vocoding.

There have been many techniques proposed in the literature to address the problem of vocoding. STRAIGHT \cite{STRAIGHT} is one of the highly successful techniques, that has proved to reduce buzzyness in the synthesized speech. However, STRAIGHT is a very high dimensional spectrum domain representation of speech and cannot be directly used in a SPSS because of the prohibitive computational cost. Therefore, in the traditional HMM-based text-to-speech (HTS) \cite{zenbc} implementations the high-dimensional spectral envelope is represented in a compressed form using Mel-general cepstrum (MGC) . In this paper we propose an alternate representation of the spectral envelope in the spectrum domain itself in a relatively lower dimension than STRAIGHT but higher than MGC. Our focus is not on achieving very low dimensionality but on using a spectral domain representation which is amenable for SPSS task. The proposed technique involves dividing the entire spectrum into sub-bands and taking only the maximum in each sub-band to represent the high-dimensional spectral envelope. During synthesis time we associate these maximum spectrum amplitudes with the centre frequencies of the respective bands and use linear/cubic interpolation to reconstruct back the entire spectrum. Our subjective and objective results indicate that spectral envelope can be reliably estimated from the short-time Fourier magnitude spectrum using the maximum spectral amplitudes in sub-bands (abbr. MSASB).

An initial investigation of using the proposed features in the SPSS framework is carried out. A deep neural network based approach is used for SPSS. Voices are synthesized using the original STRAIGHT spectral parameters as well as compressed parameters and are presented for listening tests details of which are detailed in Section \ref{sec:er2}. 

The paper is organized as follows: In the next section relation to previous work is discussed, a detailed description of the proposed analysis-by-synthesis (AbS) framework is then give in Section 3 and successively objective and subjective evaluation of the method and results are presented in Section 4. We present our experiments on using the proposed parametrization in Section 5 and its evaluations are done in Section 6. Conclusions and Future Work are discussed in the end.  

\section{Relation to Previous Work}
\label{sec:rpw}
The HTS STRAIGHT demo\footnote{http://hts.sp.nitech.ac.jp/?Download} involves extracting spectrum, aperiodicity and $F_0$ using STRAIGHT for speech parametrization. Then the spectrum is converted to a low dimensional representation using Mel-generalized cepstrum and the aperiodicities are converted to band aperiodicities using perceptually motivated bands. This work is motivated from the observation in \cite{huic14}, that maximum spectral amplitudes in sub-bands can be utilized for synthesis. However there are some important differences between the work in \cite{huic14} and proposed AbS technique.The work in \cite{huic14} is based on a sinusoidal AbS scheme and hence the main motivation for the use of sub-bands is to get a fixed dimensional representation. Also they use perceptually motivated sub-bands which are fewer than necessary to synthesize natural speech and hence make use of spectral amplitudes at band edges and dynamic coefficients. Because the number of bands in higher frequencies are sparse, they resort to adding a random noise component, so that fricatives can be synthesized well. An initial investigation for parametric speech synthesis was shown in \cite{hu2014investigation} to perform slightly better than baseline HTS. But in our work, we use homomorphic AbS \cite{avo} framework and hence there is no problem of fixed dimensional representation. Our sub-bands are of fixed bandwidth, non-overlapping and are higher in number than in \cite{huic14}, consequently we don't use any dynamic coefficients or band edge spectral components or random noise component.

 
\section{Spectral Envelope Estimation using MSASB}
\label{sec:semsasb}
In the voiced model of speech, vocal tract transfer function (VTTF) values can be obtained only at the harmonics, then the problem of recovering the full VTTF at all frequencies can be seen as an interpolation problem. The number of harmonics are not constant and keep varying, so we decided to split the spectrum into fixed number of bands and treat maximum in each sub-bands as sample of VTTF. The recovery of full VTTF from these sub-band maximum values is explained below.

\begin{figure}[ht!]
\begin{minipage}[b]{1.0\linewidth}
  \centering
  \centerline{\includegraphics[scale=0.18]{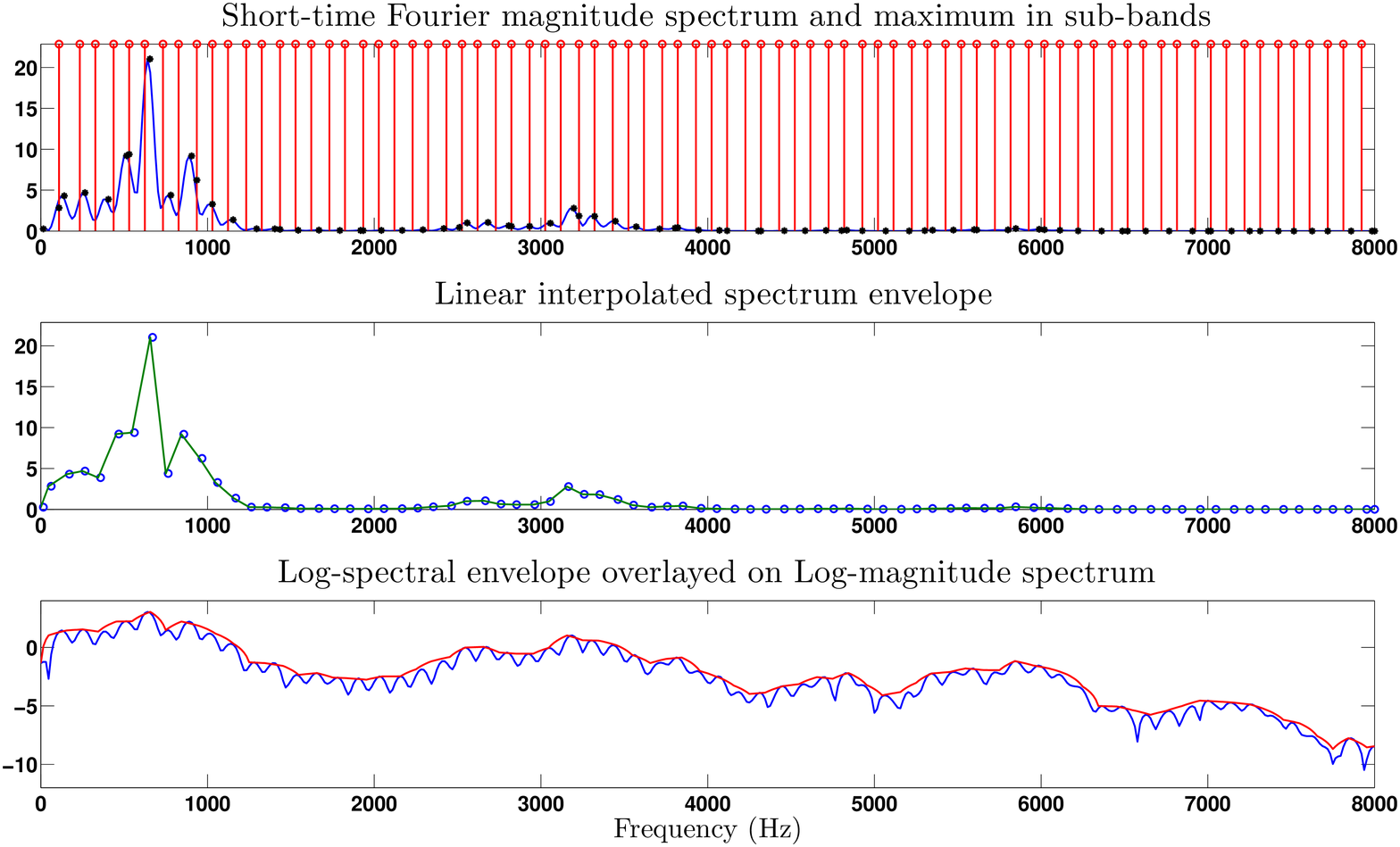}}
  \caption{MSASB on short-time Fourier magnitude spectrum }
  \label{msasbst} 
\end{minipage}
\end{figure}

The given speech signal is analyzed pitch adaptively using a Hanning window of 3 pitch periods. In unvoiced regions a constant window length of 15ms is used. The Voiced/Unvoiced decision is obtained using $F_0$ contour ( in our case STRAIGHT $F_0$ was used but any $F_0$ estimation \cite{zff} and voiced/unvoiced algorithm \cite{vuv} will work). Then the MSASB procedure is applied to extract the spectral envelope from the short-time Fourier magnitude spectrum. The procedure is depicted in Fig. \ref{msasbst}. The first step is to compute the magnitude spectrum of the windowed signal by discrete Fourier transform. Then each spectral slice/frame is split into $N_b$ sub-bands. These sub-bands are non-overlapping and of fixed bandwidth. The bandwidth of each sub-band can be found as $\frac{fs}{2*N_b}$. The maximum in each sub-band is then computed and stored, in addition the values of magnitude spectrum at $0,\frac{fs}{2}$ Hz are also stored. This is done for a constant frame shift of 5ms. This whole process can be viewed as analysis stage.  

During the synthesis stage, an impulse response must be first obtained which then is convolved using an excitation signal. The estimate of the vocal tract transfer function is obtained from the maximum spectral amplitudes in sub-bands as follows. Because we don't store the frequencies at which the maximum spectral values have resulted in, during synthesis we align these to the band centre frequencies. Although, this adjustment of spectral energies can lead to distortion, we find in our listening tests that this effect is almost negligible. While reconstruction of the full spectral envelope, these $N_b+2$ samples are interpolated using linear/cubic methods. Once this is done, a minimum phase version of spectral envelope is obtained by suitably weighting the cepstrum. The minimum phase impulse response is then generated which needs to be convolved with a excitation source signal. The synthetic source signal can be generated in many ways, simplest of which is to generate an impulse train for voiced regions and random noise in unvoiced regions. But the spectral envelope estimation technique described above can also be used with other advance forms of source generation schemes like STRAIGHT. In our AbS experiments, we use the estimated spectral envelope with STRAIGHT source parameters to synthesize speech signal. 

\begin{figure}[ht!]
\begin{minipage}[b]{1.0\linewidth}
  \centering
  \centerline{\includegraphics[scale=0.18]{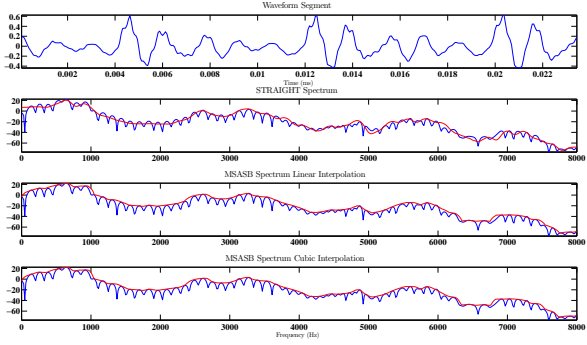}}
  \caption{MSASB on short-time Fourier magnitude spectrum Vs. STRAIGHT spectrum envelope}
  \label{msasbse} 
\end{minipage}
\end{figure}

Fig. \ref{msasbse} shows an example waveform segment and the spectral envelope from STRAIGHT, MSASB using linear and cubic methods overlayed on the magnitude spectrum. Both the interpolation schemes linear and cubic seem to give very similar looking spectrum shapes.

\section{Experiments and Results - 1}
\label{sec:er1}
In this section we present experiments related to AbS using MSASB technique. A dataset containing 2 female and 2 male speakers from ARCTIC  \cite{arctic} (BDL,RMS,SLT and CLB) was taken with 10 utterances of each speaker. Specifically we experiment on how many number of bands give a high-quality synthetic speech.
The number of bands $N_b$ was varied from 60 to 160 in steps of 20. The following AbS systems were presented for evaluation. 

\begin{itemize}
\item ST (S1) : STRAIGHT AbS system.
\item MSASB60 (S2): MSASB using 60 sub-bands.
\item MSASB80 (S3): MSASB using 80 sub-bands.
\item MSASB100 (S4): MSASB using 100 sub-bands.
\item MSASB120 (S5): MSASB using 120 sub-bands.
\item MSASB140 (S6): MSASB using 140 sub-bands.
\item MSASB160 (S7): MSASB using 160 sub-bands.
\end{itemize}

\begin{figure}[htb]
\begin{minipage}[b]{1.0\linewidth}
  \centering
  \centerline{\includegraphics[scale=0.18]{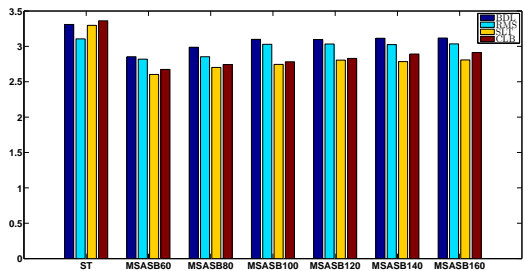}}
  \caption{ PESQ scores for AbS with STRAIGHT and MSASB with varying sub-bands}
  \label{pesqmos_nbands} 
\end{minipage}
\end{figure}

\begin{figure}[ht!]
\begin{minipage}[b]{1.0\linewidth}
  \centering
  \centerline{\includegraphics[scale=0.18]{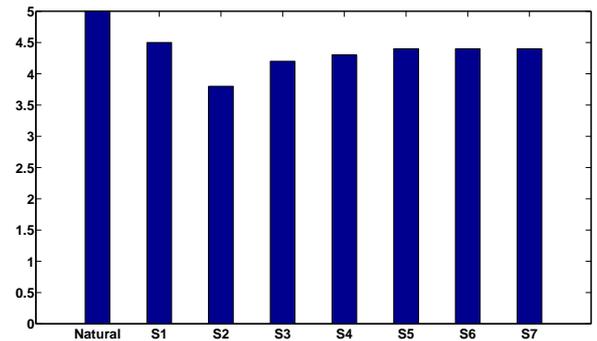}}
  \caption{ MOS scores of STRAIGHT and various MSASB schemes}
  \label{mos}
\end{minipage}
\end{figure}

We conducted a listening test containing 20 listeners. Fig. \ref{mos} shows the MOS scores of systems. ``Natural'' represents original recordings. As an objective measure, perceptual speech quality (PESQ) \cite{pesq} scores were computed and the average PESQ scores are reported in Table \ref{pesq}. The pesq scores averaged over 10 utterances of male voices are approaching that of STRAIGHT while that of female is slightly lower as can be seen from \ref{pesqmos_nbands}. However the reason for such a result is not evident and has to be investigated. As is evident from the Table \ref{pesq} decreasing the number of sub-bands affects the quality of speech more. The code for replicating the experiments can be found at \footnote{https://github.com/SivanandAchanta/MSASB}.

Clearly with 100 sub-bands for spectrum we are able to synthesize speech that is very close to STRAIGHT synthesis. This is confirmed from the listening test with as MOS score of 4.4 as high as our baseline method STRAIGHT. Although, the objective PESQ scores were slightly higher for STRAIGHT, subjective MOS scores indicate that both the methods synthesize equally well. Since it is not clear how well objective scores correlate with subjective listening experience, MOS score might be a slightly better indicator of our results. The samples can be heard at \footnote{\url{http://goo.gl/2ZTlRr}}. In view of this we experiment the proposed parametrization in DNN-SPSS framework detailed in the below section. 

\begin{table}[ht!]
\centering
\eightpt \caption{Objective PESQ scores (mean of all 4 speakers) obtained by comparing natural with the synthesized versions. Higher scores imply better AbS systems. }
\label{pesq}
\begin{tabular}{|c | c | c| c | c |c |}
  \hline
  \eightpt Method & \eightpt S1 & \eightpt S2  & \eightpt S3 & \eightpt S4 & \eightpt S7\\
  \hline
  \eightpt PESQ  & 3.27 & 2.73 & 2.82 & 2.91 & 2.96\\
  \hline 
  
\end{tabular}
\end{table}

\section{Deep Neural Network (DNN) based Statistical Parametric Speech Synthesis}
\subsection{Experiments and Results - 2}
\label{sec:er2}
Using the propose AbS scheme we built DNN based SPSS systems. Our experiments were done using  SLT and BDL, female and male speakers in ARCTIC database \cite{arctic} respectively. We have extracted the full context labels from labelled data. This includes quinphone identities along with vowel in the current syllable as categorical features. The numerical features include the number of phones in the syllable, number of syllables in the word, number of words in the utterance and so on. The total dimension of the input feature vector was 305. The phone level duration and frame indicators were included as duration features at input. Note that natural duration's were used in our experiments. Input features were mean and variance normalized. 

At the output, spectrum extracted using STRAIGHT tool was used for baseline and MSASB (with logarithm applied) values for our method. The frame shift was set to 5ms. The output features were left unnormalized (we found that normalizing between 0.01 and 0.99 did not improve the performance). We have take natural $F_0$, and aperiodicity during synthesis time. This makes sure that the observed effects are purely because of different spectral envelopes used.

The architecture of the DNN used was 305L 1050R 1050R xL, where R implies rectified linear units (ReLU \cite{relu}) and L for linear neurons. The x at output layer can be either 204 or 1026 depending on whether MSASB was used at the output or STRAIGHT spectrum is used. We use deltas alone at the output. No MLPG \cite{dnnsps} \cite{hmmspss} like smoothing was performed on the predicted spectra, rather raw predicted spectra were combined with natural $F_0$ and aperiodicity for synthesizing speech. 

There were total 1131 sentences of which 913 were used for training, 100 for validation and remaining for testing. AdaDelta \cite{adadelta} was used for learning rate setting and Nesterov's accelerated gradient based momentum \cite{sutsrnn} was used. The momentum factor was set to 0.9 after fine-tuning on validation dataset. The training was terminated after 200 epochs, as the average normalized mean squared error no longer decreased appreciably. The mini-batch size was set to 1000. We initially tried with a mini-batch size of 200, altering it to 1000 greatly increased speed but dint affect the performance so we chose to go with 1000. A GPU-based mini-batch stochastic gradient descent (M-SGD) was implemented in Matlab. The experiments were run on NVIDIA Geforce GTX-660 graphics card. Each setup took less than 1Hr to train the voice. The code for replicating the experiments is available at \footnote{https://github.com/SivanandAchanta/DNN\_SPS} and samples at \footnote{http://goo.gl/s4945L}. 

A subjective preference test was conducted by synthesizing test data, using STRAIGHT spectrum and MSASB spectrum. The results are shown in Table \ref{sps}, which indicate that the preference for both the systems is more or less equal. This means that MSASB parameters are able to synthesize as good as STRAIGHT spectrum when incorporated into an SPSS framework. 

\begin{table}[ht]
\centering
\eightpt \caption{Subjective preference scores (in \%) comparing the SPSS using baseline STRAIGHT method and MSASB\_100 (detailed in Section \ref{sec:er1})}
\label{sps}
\begin{tabular}{|c | c | c|}
  \hline
  SPK & ST & MSASB100\\
  \hline
  BDL & 55 & 45 \\
  \hline
  SLT & 52 & 48 \\  
  \hline 
\end{tabular}
\end{table}


\section{Conclusions}
In this paper we presented an alternate technique for spectral envelope estimation by parametrizing spectrum in frequency domain. Subjective and objective results for AbS indicate that our MSASB method performs similar to STRAIGHT representation at a lower dimension. We have performed initial investigation of using the proposed parametrization in DNN based SPSS. Subjective preference tests on the DNN-SPSS show that the preference for both STRAIGHT and MSASB parametric representations are more or less equal meaning our parametric representation of spectrum is suitable for SPSS task. Our future work is on building SPSS systems with more data and quantify the averaging effect of DNN model on the MSASB parameters, so that post-filtering can be applied in the frequency domain itself. 
 
\section{Acknowledgements}  
The authors would like to thank Dr. Kishore Prahallad for helpful discussions and Tejas G for proof reading the manuscript.

\eightpt
\bibliographystyle{IEEEtran}

\bibliography{msasb}


\end{document}